\documentclass[aps,prd,floatfix,superscriptaddress,preprintnumbers]{revtex4}
\usepackage{titlesec}
\usepackage{amssymb,amsmath,amsfonts,latexsym,graphicx,epsfig}
\usepackage{empheq}
\usepackage{float}
\usepackage{xcolor}

\newcommand{\bea}{\begin{eqnarray}}
\newcommand{\ena}{\end{eqnarray}}
\newcommand{\be}{\begin{equation}}
\newcommand{\en}{\end{equation}}
\newcommand{\nn}{\nonumber\\}

\newcommand{\Heff}{\mbox{${\cal H}_{\rm eff}$}}
\newcommand{\ed}{\end{document}}

\begin{document}

\title{Analysis of the nonleptonic two-body decays of the \boldmath{$\Lambda$} hyperon} 

\author{Mikhail~A.~Ivanov}
\affiliation{Bogoliubov Laboratory of Theoretical Physics,
Joint Institute for Nuclear Research, Dubna, Russia}
\author{J\"urgen~G.~K\"orner}
\affiliation{PRISMA Cluster of Excellence, Institut f\"{u}r Physik,
Johannes Gutenberg-Universit\"{a}t, D-55099 Mainz, Germany}
\author{Valery~E.~Lyubovitskij}
\affiliation{Institut f\"ur Theoretische Physik, Universit\"at T\"ubingen,
Kepler Center for Astro and Particle Physics,
Auf der Morgenstelle 14, D-72076 T\"ubingen, Germany}
\affiliation{Departamento de F\'\i sica y Centro Cient\'\i fico
Tecnol\'ogico de Valpara\'\i so-CCTVal, Universidad T\'ecnica
Federico Santa Mar\'\i a, Casilla 110-V, Valpara\'\i so, Chile} 
\affiliation{Millennium Institute for Subatomic Physics at 
the High-Energy Frontier (SAPHIR) of ANID, \\
Fern\'andez Concha 700, Santiago, Chile}
\affiliation{Department of Physics, Tomsk State University,
634050 Tomsk, Russia} 
\affiliation{Tomsk Polytechnic University, 634050 Tomsk, Russia} 
\author{Zhomart~Tyulemissov}
\affiliation{Bogoliubov Laboratory of Theoretical Physics,
Joint Institute for Nuclear Research, Dubna, Russia}
\affiliation{ The Institute of Nuclear Physics, Ministry of Energy of
the Republic of Kazakhstan, 050032 Almaty, Kazakhstan} 
\affiliation{ Al-Farabi Kazakh National University, 050040 Almaty, Kazakhstan}

\begin{abstract}

We systematically study two-body nonleptonic decays of light lambda 
hyperon $\Lambda \to p \pi^- (n\pi^0)$  
with account for both  short- and long-distance effects. 
The short-distance effects are induced 
by five topologies of external and internal weak $W^\pm$ exchange, 
while long-distance effects are saturated by an inclusion of 
the so-called pole diagrams with intermediate $\frac12^+$ and
$\frac12^-$ baryon resonances. The contributions from
$\frac12^+$~resonances are calculated straightforwardly by account
for nucleon and $\Sigma$~baryons whereas  the
contributions from $\frac12^-$~resonances are calculated by using
the well-known soft-pion theorem in the current-algebra approach.
It allows one to express the parity-violating $S$-wave amplitude in terms
of parity-conserving matrix elements. 
From our previous analysis of heavy baryons we know that 
short-distance effects induced by internal topologies are not 
suppressed in comparison with external $W$-exchange diagram 
and must be included for description of data. 
Here, in the case of $\Lambda$ decays we found that the contribution 
of  external and internal $W$-exchange diagrams is sizably suppressed, e.g., 
by one order of magnitude in comparison with data, which are 
known with quite good accuracy. Pole diagrams play the major role 
to get consistency with experiment. 

\end{abstract}

\today

\maketitle

\section{Introduction}
\label{sec:intro}

Two-body nonleptonic decays of the $\Lambda$ hyperon $\Lambda \to p \pi^-$ and 
$\Lambda \to n \pi^0$ are the leading modes with branching fractions 
$63.9 \pm 0.5\%$ and $35.8 \pm 0.5\%$~\cite{PDG20}, respectively.
These processes follow by producing $\Lambda$ in the $K^- p$ scattering
and are clearly  identified after substraction of background effects. 
The other modes of the $\Lambda$ are suppressed by a few orders of magnitude.
Such precise data on the $\Lambda \to N \pi$  decays give a unique possibility
to consider these decays as a laboratory for testing 
nature of the weak decays and a crucial check of existing theoretical
approaches modeling exclusive decays of baryons.
From a modern theoretical point of view, there are two classes of
the Feynman diagrams  generating matrix elements of these processes:
(1) short-distance (SD) diagrams and 
(2) long-distance (LD) or pole diagrams. 
The SD diagrams are classified by five different color-flavor topologies
as shown in Fig.~\ref{fig:tpg}. We refer to the topologies of Ia
and Ib as tree diagrams. They are sometimes also
referred to as external (Ia) and internal $W$-emission (Ib) diagrams.
The topologies IIa, IIb, and III are referred to as $W$--exchange diagrams.
Note, in literature one can find other notations for $W$--exchange diagrams. 
For example, in~\cite{Leibovich:2003tw} they are denoted as the exchange (IIa),
color-commensurate (IIb) and Bow tie (III) diagram. As shown in
Fig.~\ref{fig:tpg} the color-flavor factor of the tree diagrams Ia and Ib
depends on whether the emitted meson is charged or neutral. For charged
emission the color-flavor factor is given by the linear combination
of Wilson coefficients $(C_2 + \xi C_1)$, where
$\xi=1/N_c$, while for neutral emission the color-flavor factor reads
$(C_1 + \xi C_2)$. 
We use the large--$N_c$ limit for the color-flavor factors with $\xi=0$. 
For the $W$--exchange diagrams  the color-flavor factor is
given by $(C_2-C_1)$. The Wilson coefficients $C_2=1.361$ (leading order)
and $C_1=-0.625$ (subleading order) are taken at the scale $\mu=1$~GeV,
on $\Lambda^{(4)}_{\overline \rm MS}=435$~MeV 
in the Naive Dimensional Regularization (NDR) scheme
from Ref.~\cite{Buchalla:1995vs}. 
Note, the contribution of the QCD penguin diagrams to the nonleptonic decays of
the $\Lambda^0$ hyperon is strongly suppressed in comparison with
the contribution of the current-current operators due to a suppression of 
the corresponding Wilson coefficients. 

In Refs.~\cite{Ivanov:2020xmw,Ivanov:2020iaq,Gutsche:2019iac,Gutsche:2019wgu,%
    Gutsche:2018msz,Gutsche:2018utw,Gutsche:2017hux,Gutsche:2017wag,%
  Gutsche:2015lea,Ivanov:1997ra,Ivanov:1997hi}   
we have performed  comprehensive analysis of the nonleptonic decays of
the single and  double heavy baryons including topologies with external and 
internal $W$ exchange. In particular, we have shown that the internal
topologies are  not always suppressed and must be systematically
included in theoretical  analysis. Here we turn to study of nonleptonic decays
of the light baryons, which up to now were more precisely determinated from
the data and could give an opportunity  for an additional check of the
theoretical approaches. 

Since the end of the 1960s two-body nonleptonic decays of the $\Lambda$ hyperon 
have been studied in the literature using different approaches. 
One of the first attempts to the $\Lambda$ and other hyperons 
has been done by using effective weak Hamiltonians 
and methods of current algebra, dispersion relations, and vector 
dominance~\cite{Sakurai:1967zzc,Balachandran:1967zz,Nussinov:1968xh,Chan:1968qp,
  Albright:1969xu,Hosoya:1971gq,Nagai:1972mk,Scadron:1973mv,Sebastian:1974np,
  Khanna:1975vt}. In Ref.~\cite{Nakagawa:1967hn}
nonleptonic decays of hyperons were analyzed using 
three types of nonrelativistic bound-state models of baryons: a quark model, 
a three-triplet model and a quartet model. In the 1970s significant progress was
achieved in the studying nonleptonic decays of hadrons (e.g., kaons
and hyperons). It was shown (for review, see Ref.~\cite{Shifman:1976ge})
that short-distance strong  effects playing important role in weak interactions
of hadrons can be systematically included in QCD, which lead to modification
of effective weak Hamiltonians by dressing their couplings and arising of
new operators beyond a $V-A$ structure.  In this vein, effective weak
Hamiltonians~(for review see, e.g., Ref.~\cite{Buchalla:1995vs}) 
are constructed with further their use in evaluation of matrix elements to 
predict nonleptonic decay properties of hadrons. First systematic application
of these ideas to hyperons and also to kaons can be done in
Ref.~\cite{Shifman:1976ge}.  In particular, in Ref.~\cite{Shifman:1976ge}
the most general form of the effective 
weak Hamiltonian dressed by strong short-distance effects and 
containing four-quark operators with specific 
quantum numbers of isospin and unitary spin has been derived. Dressing 
of the couplings in operators due to gluon corrections has been taking into 
account by solving corresponding renormalization group equations. 
Finally, matrix elements of nonleptonic transitions have been evaluated using 
the valence quark approximation. This work~\cite{Shifman:1976ge} served 
as the basis for modern theory of weak decays of hadrons. The effective weak
Hamiltonian is a universal ingredient in calculation of matrix elements of
physical processes and all model ambiguities are encoded in a way of
projection of the Hamiltonian between respective hadronic states or
to evaluation of matrix elements.  Note, that significant progress in studying
nonleptonic decays of light baryons  has been made in the framework of the
effective field theories~\cite{Lee:1968ehl,Bijnens:1985kj,Jenkins:1991bt,%
  Carone:1991ni,Borasoy:1999md,Borasoy:1998ku,Springer:1999sv,AbdElHady:1999mj,%
  Flores-Mendieta:2019lao,Tandean:2003fr}.
In particular, different types of effective theories [chiral perturbation 
theory (ChPT], heavy baryon ChPT (HBChPT), and large--$N_c$ QCD]  with 
implementation of chiral and $1/N_c$ corrections, isospin, and $SU(3)$ 
breaking correction for study of hyperon decays have been developed. 
It was found that an extension of ChPT beyond leading order is extremely
important and  one-loop/chiral corrections became significant for
the $p$-wave hyperon amplitudes.  Recently, an updated analysis of light baryon
nonleptonic decays in ChPT has been performed 
in Ref.~\cite{Flores-Mendieta:2019lao} with taking into account large--$N_c$
expansion and  $SU(3)$ symmetry-breaking effects. In Ref.~\cite{Tandean:2003fr} 
the contributions to the $CP$-violating asymmetries 
induced by the chromomagnetic-penguin operators have been studied. 
In Ref.~\cite{Nardulli:1987ns} weak and radiative decays of hyperons
have been considered in a pole model. In Ref.~\cite{Wu:1985yb} nonleptonic
decays of hyperons were analyzed in a constituent quark model. Skyrme soliton
model was applied for a description of weak nonleptonic decays of hyperons in 
Refs.~\cite{Donoghue:1985dya,Praszalowicz:1985gz}. 
The MIT bag model to nonleptonic decays of baryons has been applied in 
Ref.~\cite{Galic:1979tw}. Phenomenological analysis of nonleptonic 
decays of hyperons-based chiral Lagrangian model 
was done in Ref.~\cite{Pervushin:1984ik}. 
Sum rules for the nonleptonic hyperon decays 
have been derived in Refs.~\cite{Schmid:1976pt,Katuya:1978ps,%
Terasaki:1976mz,Guberina:1977ct}.  
In Ref.~\cite{LeYaouanc:1977ys} nonleptonic decays have been considered 
in nonrelativistic potential model. 
In Ref.~\cite{Hayashi:1978he} nonleptonic decays of hyperons 
have been analyzed in quark-diquark model. 
$SU(4)$--flavor chiral soliton model was applied to a description of 
hyperon nonleptonic decays in Ref.~\cite{Abe:1986xb};  
the two-body nonleptonic decays of light baryon octet and decuplet have been 
studied using combination of the topological diagram approach and 
the $SU(3)$ irreducible representation approach, i.e., the transition 
amplitudes are derived as linear combinations of the basic amplitudes classified 
accordingly to irreducible representations of the unitary flavor group 
and constraints due to isospin symmetry we also implemented.
The main conclusion of Ref.~\cite{Xu:2020jfr} is that $W$-exchange diagrams
give large and even dominant contributions to the decay rates.
Note that in these papers, a nontrivial nonperturbative dynamics of quarks
in hadrons and in the intermediated stage in propagating between individual
hadrons was modelled by the dipolelike form factors~\cite{Wang:2019alu} or
even ignored as in Ref.~\cite{Xu:2020jfr}. 
From our analysis of nonleptonic decays we know that it is a sharp
approximation,  which not always matches data. In particular, one can
describe data up to order of magnitude,  but not precise. A modeling of the
internal nonperturbative dynamics of hadronic constituents is needed 
in study of exclusive decays of baryons. 
In Ref.~\cite{Berdnikov:2007zza} two-body nonleptonic decays of the
$\Lambda$ have been considered in  an effective quark model with chiral
$U(3) \times U(3)$ symmetry. Partial decay rates and 
angular distributions have been calculated. Based on $SU(2)$ spin, $SU(3)$
flavor symmetries and vector dominance joined description of weak radiative
and nonleptonic decays of light hyperons has been done in
Ref.~\cite{Zenczykowski:2005cs}. In Ref.~\cite{Cristoforetti:2004rr} 
factorizing and pole contributions to the  nonleptonic decays of light hyperons
have been evaluated taking into account 
the instanton contributions in the framework of the Random Instanton Liquid
Model.  It was concluded, that roughly 70\% of the amplitudes come from
instanton-induced interactions (responsible for the spontaneous breaking of
chiral symmetry), 10\% from hard gluon-exchange corrections, 
and the remaining 20\% were due to confinement effect. A very
weak dependence on a choice of current quark mass was noticed. 
In Ref.~\cite{Hiyama:2004wd}
a potential model explicitly incorporating quark-quark 
correlations was applied to nonleptonic decays of light hyperons. 
In Ref.~\cite{Garcia:1998aq} an algebraic approach based on mixing hyperons
was used for calculation of their nonleptonic decay rates. 

The main objective of the present paper is to investigate a role of $W$ exchange 
diagrams in strange hyperon physics. It is well known that 
the factorizable diagrams dominated by SD effects ~\cite{Okun,Galic:1979tw,Guberina:1977ct}, 
in principle, describe data well. However, it does not mean that 
$W$-exchange and LD diagrams vanish. They could nontrivially
interplay with SD factorizable contributions. 
Nonfactorizable diagrams in baryon nonleptonic decays play an important role.
The analysis of nonleptonic baryon decays is complicated by the necessity of having
to include nonfactorizing contributions. One thus has to go beyond the factorization
approximation which had proved quite useful in the analysis of the exclusive nonleptonic
decays of heavy mesons. There have been some theoretical attempts to
analyze nonleptonic heavy baryon decays using factorizing contributions alone, 
the argument being that $W$-exchange contributions can be neglected in analogy
to the power-suppressed $W$-exchange contributions in the inclusive nonleptonic decays
of heavy baryons. One might even be tempted to drop the nonfactorizing contributions
on account of the fact that they are superficially proportional to $1/N_c$.
However, since $N_c$-baryons contain $N_c$ quarks an extra combinatorial factor proportional
to $N_c$ appears in the amplitudes which cancels the explicit diagrammatic $1/N_c$ factor.
Another argument supporting importance of study of $W$-exchange diagrams is that
there are the modes which are nonsuppressed and proceed only via these diagrams.
(see detailed discussion in Refs.~\cite{Korner:1992wi,Ivanov:1997ra}). 
In Ref.~\cite{Ivanov:1997ra} we showed
that the total contribution of the nonfactorizing diagrams can amount up to 60\% of
the factorizing contribution for heavy-to-light baryon transitions and up to 30\% for
$b \to c$ baryon transitions. Recently we improved our formalism for study of the
nonfactorizable diagrams in the nonleptonic decays of heavy baryons. 
We have made an {\it ab initio} 
three-loop quark model calculation of the $W$-exchange contribution
to the nonleptonic two-body decays of the doubly charmed baryons $\Xi^{++}_{cc}$ 
and $\Omega^+_{cc}$. 
The $W$-exchange contributions appear in addition to the factorizable tree graph contributions
and are not suppressed in general. We made use of the covariant confined quark model previously
developed by us to calculate the tree graph as well as the $W$-exchange contribution.
We calculated helicity amplitudes and quantitatively compare the tree graph and $W$-exchange
contributions. Finally, we compared the calculated decay widths with those from other
theoretical approaches when they are available. 
We found a substantial contribution of $W$-exchange graphs to the modes
with final baryon containing spin-0 light diquarks. The suppression of the
$W$-graphs for the modes with final baryons containing spin-1 light diquarks
is explained by the consistency of our framework with the K\"orner,Pati, Woo (KPW)
theorem~\cite{Korner:1970xq,Pati:1970fg}, which states that the contraction 
of the flavor-antisymmetric current-current
operator with a flavor-symmetric final-state configuration is zero in the $SU(3)$ limit. 
In Ref.~\cite{Gutsche:2019iac} we made unified analysis of semileptonic and 
nonleptonic two-body decays 
of the double-charm baryon ground states $\Xi^{++}_{cc}$, 
$\Xi^+_{cc}$, and $\Omega^+_{cc}$.
We identified those nonleptonic decay channels in which the decay proceeds
solely via the factorizing contribution precluding a contamination from $W$-exchange.
We use the covariant confined quark model previously developed by us to calculate
the various helicity amplitudes which describe the dynamics of the 
$1/2^+ \to 1/2^+$ and $1/2^+ \to 3/2^+$ 
transitions induced by the Cabibbo favored effective ($c \to s$) and ($d \to u$) currents.
We then proceed to calculate the rates of the decays as well as polarization effects
and angular decay distributions of the prominent decay chains resulting from
the nonleptonic decays of the double heavy charm baryon parent states.
Taking into account above arguments we conclude that taking into account of $W$-exchange graphs 
and development of theoretical methods for their evaluation is quite an important task. 

The paper is structured as follows.
In Sec.~\ref{sec:approach} we give the basic ingredients  of our framework
which includes the effective Hamiltonian of the weak interactions and
the description of quark structure of hadrons in the covariant
confined quark model.
Sec.~\ref{sec:analitic} is devoted to calculation of the matrix elements
of the decays $\Lambda\to p+\pi^-$ and  $\Lambda\to n+\pi^0$.
We discuss in detail the classification of the Feynman diagrams appearing
in these decays and give the analytical expressions for matrix elements
and helicity amplitudes.
In Sec.~\ref{sec:results} we present numerical results for comprehensive
analysis of nonleptonic decays of $\Lambda$ hyperon. 
Finally, in Sec.~\ref{sec:summary} we make conclusions and
summarize the main results obtained in this paper.

\section{Effective Hamiltonian and diagrams}
\label{sec:approach}

   We concentrate our discussion on two nonleptonic decay modes of
   $\Lambda-$~hyperon:  $\Lambda^0 \to p + \pi^-$ and $\Lambda^0 \to n + \pi^0$.
   They proceed due to weak interactions of quarks which are described
   by the effective Hamiltonian:
\be
   \Heff = -\frac{G_F}{\sqrt{2}}\,V_{CKM}\,
   \{  C_2\,(\bar u_a O_L s_a) (\bar d_b O_L u_b)
   + C_1\,(\bar u_a O_L s_b) (\bar d_b O_L u_a) \}
\label{eq:Heff}   
\en
where $V_{CKM}=V_{\rm ud} V^\ast_{\rm us}$ is the product of
the Cabibbo-Kobayashi-Maskawa matrix elements,
the matrix $O^\mu_L = \gamma^\mu (I-\gamma_5)$ is the weak matrix
with the left chirality. The summation over $\mu$ is implied in
Eq.~(\ref{eq:Heff}). The Wilson coefficients $C_2$ (leading order)
and $C_1$ (subleading order) are taken at the scale $\mu=1$~GeV,
on $\Lambda^{(4)}_{\overline \rm MS}=435$~MeV in the NDR scheme
from Ref.~\cite{Buchalla:1995vs} (see Table~XVIII): 
$C_1 = -0.625, \ C_2  = 1.361.$

We should stress that the contribution of the QCD penguin diagrams 
to the nonleptonic decays of
the $\Lambda^0$ hyperon is strongly suppressed in comparison with
the contribution of the current-current operators because
the Wilson coefficients $C_3-C_6$ corresponding to the
the penguin operators are suppressed at least by two orders of magnitude 
in comparison with the Wilson coefficients $C_1$ and $C_2$. We 
present a comparison of the Wilson coefficients $C_1,C_2$ for the
current-current operators and $C_3-C_6$ for the QCD penguins
in Table~\ref{Wilson_Coefficients}. 

\begin{table}[H]
\caption{Wilson coefficients} 
\centering
 \begin{tabular}{cccccc}
\hline
  $C_1$    &  $C_2$    &  $C_3$     &  $C_4$    & $C_5$  &  $C_6$   \\
\hline
 $-0.2632$ & $1.0111$ &  $-0.0055$ & $-0.0806$ & 0.0004 & 0.0009   \\
\hline
\label{Wilson_Coefficients}
 \end{tabular}
\end{table}

\begin{figure}[H]
  \begin{center}
\epsfig{figure=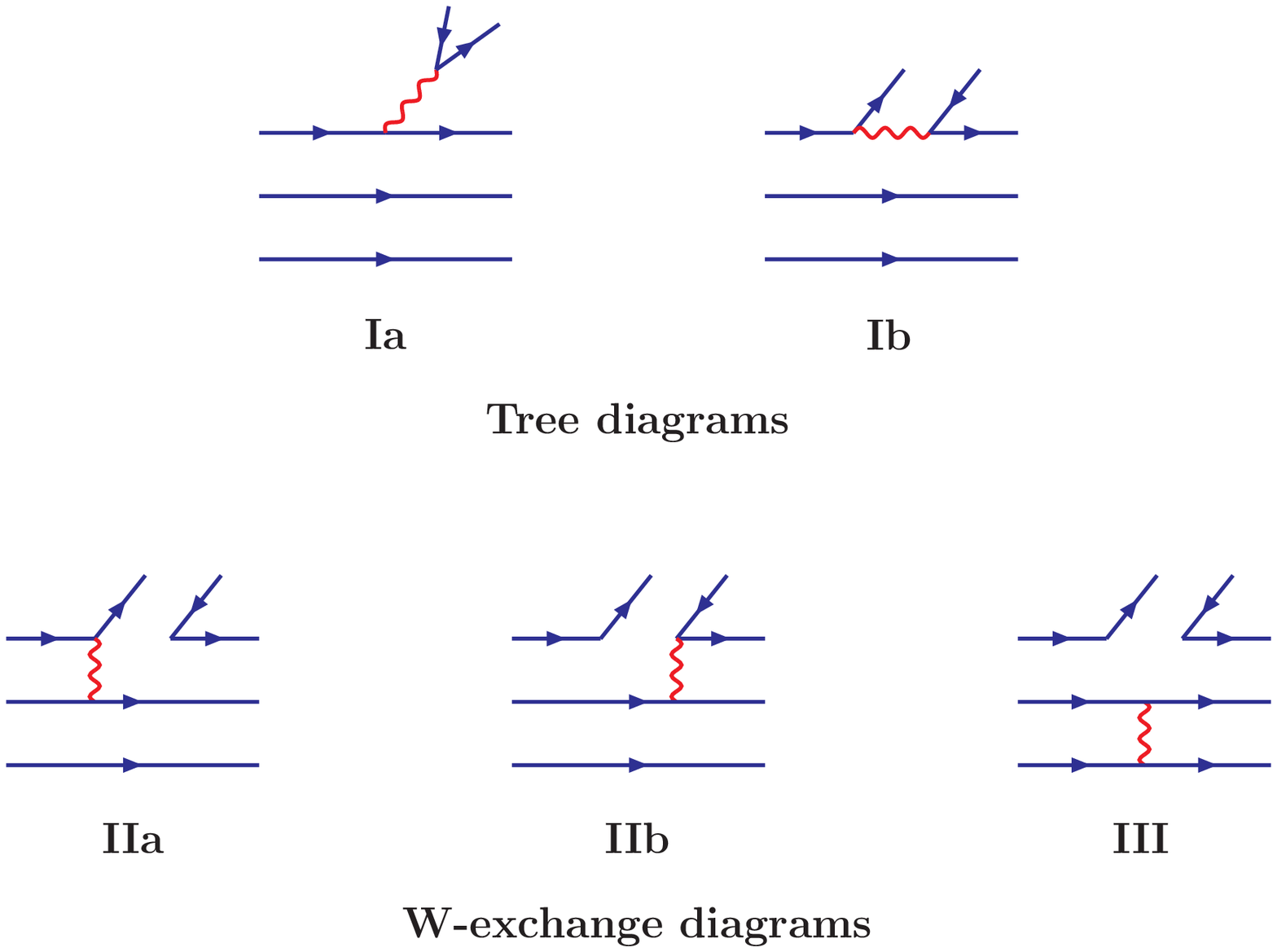,scale=.525}
\caption{Flavor-color topologies of nonleptonic weak decays.}
\label{fig:tpg}
\end{center}
\end{figure}

We will take into account both the SD-diagrams
and LD diagrams contributions. In general, the SD diagrams
have five different topologies generated by the $W$-exchange between two
quarks as shown in Fig.~\ref{fig:tpg}. 
Note, 
the $W$-exchange diagram III contributes to the $S$-wave amplitude, because
the light diquark loop with weak nonleptonic vertex contains the Dirac structure
proportional four-dimensional Levi-Cevita tensor $\epsilon^{\alpha,\beta,\rho,\sigma}$, 
which produces $S$-wave amplitude due to 
contracting with to the fermion line containing two quark propagators
and $\gamma^5$ Dirac matrix (due to pion coupling to $u/d$ quarks). 
One should stress that the contribution of the diagram III 
to the $S$-wave amplitude could vanish in some limiting cases. 
For example, in Refs.~\cite{Korner:1992wi,Ivanov:1997ra} it occurs when 
a static approximation for baryon wave functions or light quark propagators 
is used. In the present paper we go beyond static approximation. 

\begin{figure}[H]
\begin{center}
\epsfig{figure=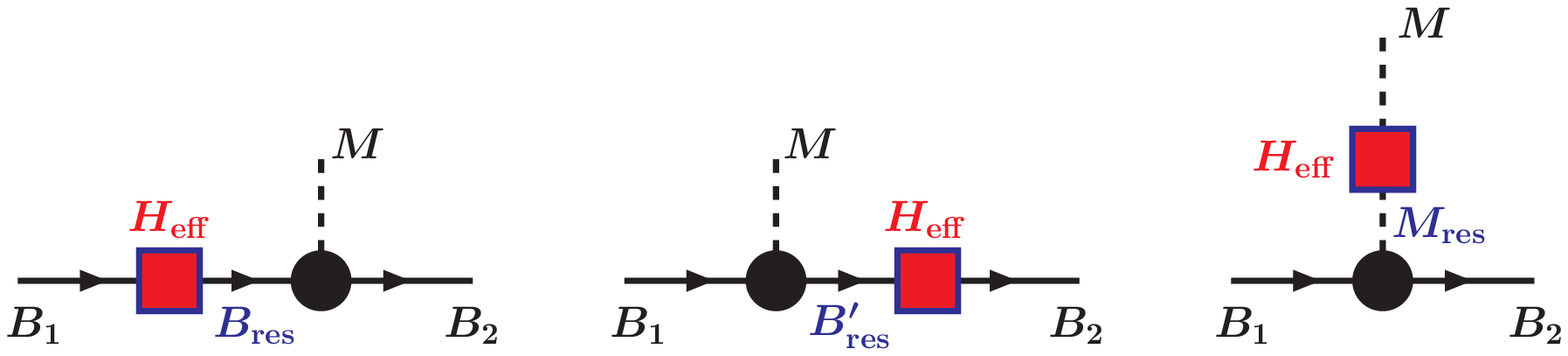,scale=.525}
\caption{The pole diagrams which effectively account for
  the long-distance contributions.}
\label{fig:pole}
\end{center}
\end{figure}

In addition to the SD diagrams we calculate the pole diagrams
as shown in  Fig.~\ref{fig:pole} by account for the lowest-lying
resonances with spin 1/2 and 0.

In Table~\ref{tab:cur} we display the quantum numbers, mass values and
interpolating currents of baryons needed in  this paper.  Note that
the forms of interpolating currents are not unique. The different
form may be transformed to each other by using the Fierz transformations.
We do not display the overall flavor factors because they will be recovered
due to $T$--product operation of the quark fields.
\begin{table}[H]
\caption{Quantum numbers and interpolating currents of baryons } 
\centering
\def\arraystretch{1.1}
\begin{tabular}{l|c|c|c}
\hline
Baryon\qquad &\quad $J^P$ \quad & \quad Interpolating current \qquad &
\quad Mass (MeV)
\\\hline
$\Lambda^0$\quad \quad \quad & $\frac12^+$ &
$\varepsilon_{abc}\, s^a (u^b C\gamma_5 d^c)$  &
1115.683 $\pm$ 0.006 
\\\hline
$\Sigma^{+}$\quad \quad \quad & $\frac12^+$ &
$\varepsilon_{abc}\,\gamma_\mu\gamma_5 \, s^a (u^b C\gamma^\mu u^c)$ &
1189.37 $\pm$ 0.07 
\\\hline
$\Sigma^{0}$\quad \quad \quad & $\frac12^+$ &
$\varepsilon_{abc}\,\gamma_\mu\gamma_5 \, s^a (u^b C\gamma^\mu d^c)$ &
1192.642 $\pm$ 0.024
\\\hline
$\Delta^{0}$\quad \quad \quad & $\frac32^+$ &
$\varepsilon_{abc}\,\{  u^a (d^b C\gamma^\mu d^c)$
-$\frac{i}{2}\gamma_\nu \, u^a (d^b C\sigma^{\mu\nu} d^c)\}$  &
1231.3 $\pm$ 0.6
\\\hline
$\Sigma^{\ast\,+}$\quad \quad \quad & $\frac32^+$ &
$\varepsilon_{abc}\,\{  s^a (u^b C\gamma^\mu u^c)$
-$\frac{i}{2}\gamma_\nu \, s^a (u^b C\sigma^{\mu\nu} u^c)\}$  &
1382.80 $\pm$ 0.35   
\\\hline
$\Sigma^{\ast\,0}$\quad \quad \quad & $\frac32^+$ &
$\varepsilon_{abc}\,\{  s^a (u^b C\gamma^\mu d^c)$
-$\frac{i}{2}\gamma_\nu \, s^a (u^b C\sigma^{\mu\nu} d^c)\}$  &
1383.7 $\pm$ 1.0      
\\\hline
$p$ \quad \quad \quad & $\frac12^+$ &
$\varepsilon_{abc}\,\{ (1-x) \gamma_\mu\gamma_5 d^a (u^b C\gamma^\mu u^c)
-\frac{x}{2}\sigma_{\mu\nu}\gamma_5 \, d^a (u^b C\sigma^{\mu\nu} u^c)\} $  &
938.2720813 $\pm$ 0.0000058         
\\\hline
$n$ \quad \quad \quad & $\frac12^+$ &
$\varepsilon_{abc}\,\{ (1-x) \gamma_\mu\gamma_5 u^a (d^b C\gamma^\mu d^c)
-\frac{x}{2}\sigma_{\mu\nu}\gamma_5 \, u^a (d^b C\sigma^{\mu\nu} d^c)\} $  &
939.5654133 $\pm$ 0.0000058
\\\hline
\end{tabular} 
\label{tab:cur}
\end{table}

We are going to calculate the matrix elements of the above-mentioned
nonleptonic decays in the framework of the covariant confined quark
model (CCQM) developed in our previous papers.
The starting point is the Lagrangian describing couplings
of the baryon field with its interpolating quark current.
\be
   {\cal L}(x) = g_B \Big( \bar B(x)\cdot J(x) + \bar J(x)\cdot B(x) \Big)
\label{eq:Lag}
\en   
where $\bar J = J^\dagger \gamma^0 $ is the Dirac conjugate current.
The coupling constant $g_B$ is determined from the normalization
condition called {\it compositeness condition}.

The nonlocal version of the interpolating currents shown in
Table~\ref{tab:cur} reads
\bea
J_B(x)  &=& \int\!\! dx_1 \!\! \int\!\! dx_2 \!\! \int\!\! dx_3 \,
F_B(x;x_1,x_2,x_3) \,
\varepsilon_{abc}\,\Gamma_1 q_1^a(x_1)\,
\left(q_2^b(x_2) \,C\Gamma_2\, q_3^c(x_3)\right)\,,
\nn
F_B &=& \delta^{(4)}\Big(x-\sum\limits_{i=1}^3 w_i x_i\Big)
\Phi_B\Big(\sum\limits_{i<j}(x_i-x_j)^2\Big) \,,
\label{eq:nonlocal-cur}
\ena
where $w_i=m_i/(\sum\limits_{j=1}^3 m_j)$ and
$m_i$ is the quark mass at the space-time point $x_i$, and $\Gamma_1,\Gamma_2$
are the Dirac strings of the initial and final baryon states as specified
in Table~\ref{tab:cur}. For simplicity and calculational advantages we mostly
adopt a Gaussian form for the  functions $\Phi_B$, i.e. we write
\bea
&&
\Phi_B\Big(\sum\limits_{i<j}(x_i-x_j)^2\Big) =
\int\!\frac{dq_1}{(2\pi)^4}\int\!\frac{dq_2}{(2\pi)^4}
e^{-iq_1(x_1-x_3)-iq_2(x_2-x_3)}
\widetilde\Phi_B\Big(-\vec\Omega^2\Big)\,,
\nn
&&
\widetilde\Phi_B\Big(-\vec\Omega^2\Big) =
\exp\left(\vec\Omega^2/\Lambda_B^2\right), \qquad
\vec\Omega^2=\tfrac12 (q_1+q_2)^2 + \tfrac16 (q_1-q_2)^2 =
\frac23\sum\limits_{i\le j} q_iq_j\,.
\label{eq:vertex}
\ena
where $\Lambda_B$ is the size parameter for a given
baryon with values of the order of 1 GeV. The size parameter 
represents the extension of the distribution of the constituent quarks
in the given baryon. 

\section{Matrix elements, helicity amplitudes and rate expressions}
\label{sec:analitic}

The matrix element of the decay  $\frac12^+ \to \frac12^+ + 0^-$ is written as
\bea
M(B_1\to B_2 + M) &=& M_{\rm SD}(B_1\to B_2 + M)
\label{eq:matr}\\[1.2ex]
&+& M_{\rm LD_1}(B_1 \to B_{\rm res}\to B_2 + M) 
+ M_{\rm LD_2}(B_1\to B'_{\rm res} + M \to B_2 +M)\,.
\nonumber
\ena
Here
\bea
M_{\rm SD} &=& i^4 \bar u(p_2)\, \Gamma_{B_1B_2M}(p_1,p_2,q)\, u(p_1)\,,
\label{eq:SD}\\
 \Gamma_{B_1B_2M} &=& g_{B_1}  g_{B_2}  g_{M} 
\int\!\! dx e^{-ip_1x}\!\!  \int\!\!dy e^{ip_2y}\!\!
\int\!\! dv e^{iqv}\!\!  \int\!\! dz\,\,
<T\Big\{ J_{B_2}(y) \Heff(z) J_M(v) \bar J_{B_1}(x) \Big\}>_{_0}\,,
\nn[1.3ex]
M_{\rm LD_1} &=&  i^6 \int\!\!\frac{d^4k}{(2\pi)^4i}\,
\bar u(p_2)\, \Gamma_{B_{\rm res} M B_2}(k,p_2,q)\, S_{B_{\rm res}}(k)\,
\Gamma_{B_1B_{\rm res} }(p_1,k)\,u(p_1)\,,
\label{eq:LD1}\\
\Gamma_{B_{\rm res} M B_2} &=& g_{B_{\rm res}}   g_{M} g_{B_2} 
\int\!\!d\xi_2 e^{-ik\xi_2}\!\! \int\!\!dy e^{ip_2y}\!\! \int\!\! dv e^{iqv}\,
<T\Big\{ J_{B_2}(y) J_M(v) \bar J_{B_{\rm res}}(\xi_2) \Big\}>_{_0}
\nn
\Gamma_{B_1B_{\rm res} } &=& g_{B_1} g_{B_{\rm res}}
  \int\!\! dx e^{-ip_1x}\!\!  \int\!\! d\xi_1 e^{ik\xi_1}\!\! \int\!\! dz\,
 <T\Big\{ J_{B_{\rm res}}(\xi_1) \Heff(z) \bar J_{B_1}(x) \Big\}>_{_0}  
\nn[1.3ex] 
M_{\rm LD_2} &=&  i^6
\int\!\!\frac{d^4k}{(2\pi)^4i}\,
\bar u(p_2)\, \Gamma_{B_{\rm res} B_2}(k,p_2)\, S_{B_{\rm res}}(k)\,
\Gamma_{B_1 M B_{\rm res} }(p_1,k,q)\,u(p_1)\,,
\label{eq:LD2}\\
\Gamma_{B_{\rm res} B_2} &=&  g_{B_{\rm res}} g_{B_2} 
\int\!\! d\xi_2 e^{-ik\xi_2}\!\!  \int\!\! dy e^{ip_2y}\!\! \int\!\! dz\,
 <T\Big\{ J_{B_2}(y) \Heff(z) \bar J_{B_{\rm res}}(\xi_2) \Big\}>_{_0}  
\nn
\Gamma_{B_1 M B_{\rm res}} &=& g_{B_1} g_{M} g_{B_{\rm res}} \,
\int\!\!dx e^{-ip_1x}\!\! \int\!\! dv e^{iqv} \!\!  \int\!\!d\xi_1 e^{ik\xi_1}\,
<T\Big\{ J_{B_{\rm res}}(\xi_1) J_M(v) \bar J_{B_1}(x)  \Big\}>_{_0}
\nonumber
 \ena 

\begin{figure}[H]
  \begin{center}
\epsfig{figure=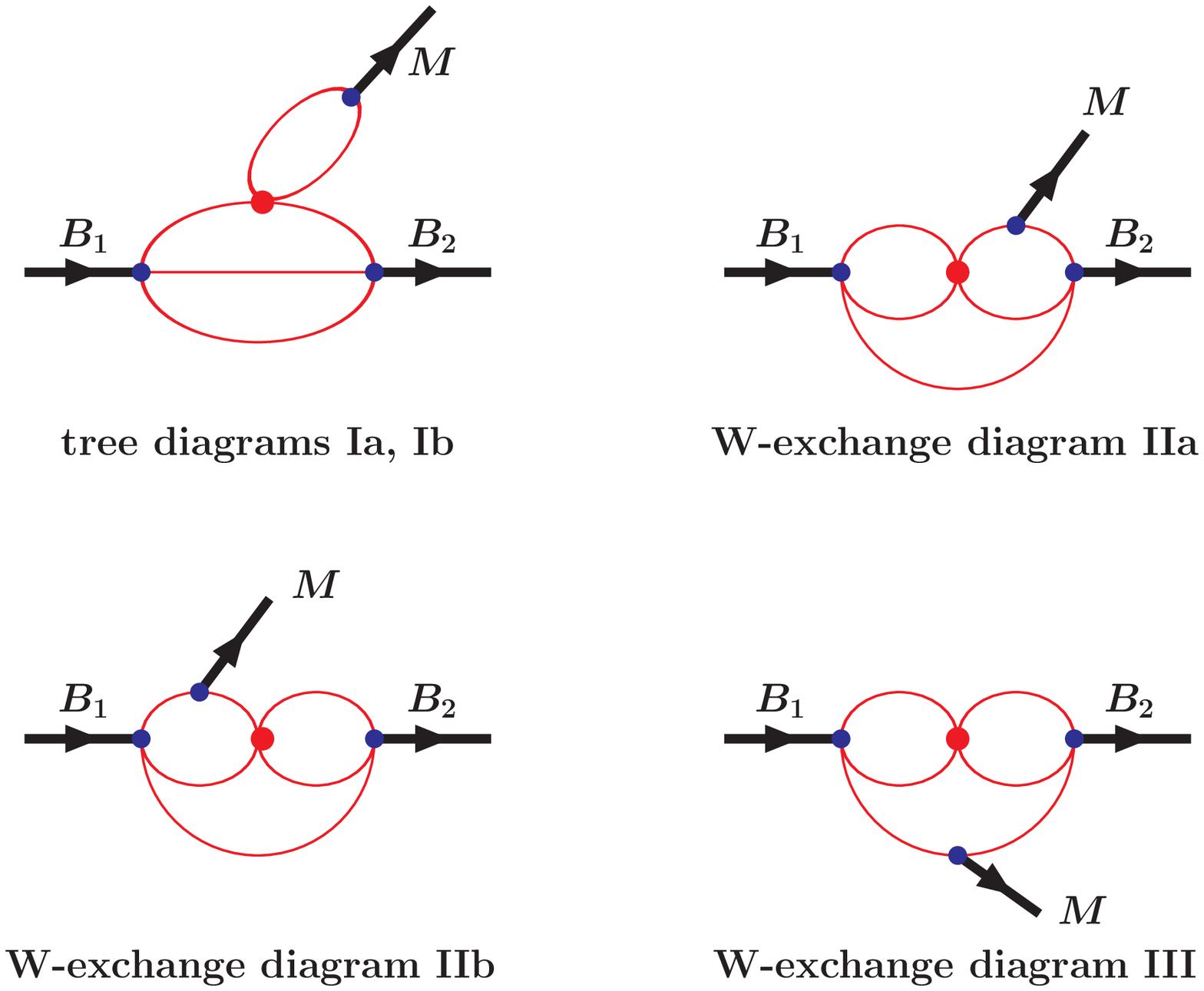,scale=.45}
\caption{Feynman diagrams describing the SD contributions.}
\label{fig:SD}

\epsfig{figure=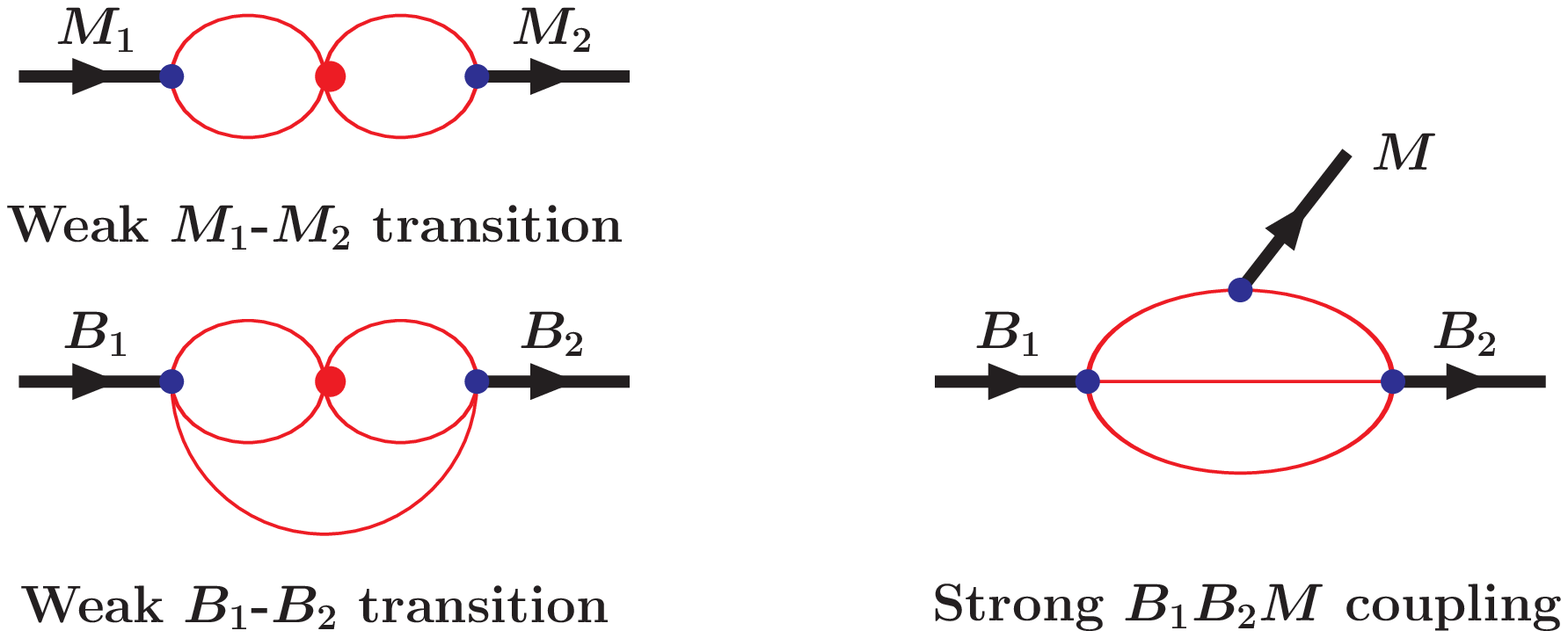,scale=.45}
\caption{Feynman diagrams describing the  building blocks of pole
  contributions.}
\label{fig:LD}
\end{center}
\end{figure}

 In our approach the building blocks $\Gamma$ are represented by
 a set of the Feynman diagrams shown in Fig.~\ref{fig:SD} for the
 SD contributions and in Fig.~\ref{fig:LD} for the
 LD contributions.

By going to the momentum space in the above building blocks one gets
\bea
\Gamma_{B_1B_2M}(p_1,p_2,q) &=& i(2\pi)^4 \delta(p_1-p_2-q)\,
\frac{G_F}{\sqrt{2}} V_{CKM}\, \widetilde\Gamma_{B_1B_2M}(p_1,p_2)\,,
\nonumber\\[1mm]
\Gamma_{B_1B_{\rm res} }(p_1,k) &=& i(2\pi)^4 \delta(p_1-k)\,
\frac{G_F}{\sqrt{2}} V_{CKM}\, \widetilde\Gamma_{B_1B_{\rm res} }(p_1)\,,
\nonumber\\[1mm]
\Gamma_{B_{\rm res} M B_2}(k,p_2,q) &=& i(2\pi)^4 \delta(k-p_2-q)\,
\frac{G_F}{\sqrt{2}} V_{CKM}\, \widetilde\Gamma_{B_{\rm res} M B_2}(k,p_2)\,,
\nonumber\\[1mm]
\Gamma_{B_{\rm res} B_2}(k,p_2) &=& i(2\pi)^4 \delta(k-p_2)\,
\frac{G_F}{\sqrt{2}} V_{CKM}\, \widetilde\Gamma_{B_{\rm res} B_2}(p_2)\,,
\nonumber\\[1mm]
\Gamma_{B_1 M B_{\rm res}}(p_1,k,q) &=&  i(2\pi)^4 \delta(p_1-k-q)\,
\frac{G_F}{\sqrt{2}} V_{CKM}\, \widetilde\Gamma_{B_1 M B_{\rm res}}(p_1,k)
\label{eq:block}
\ena
where $ V_{CKM} = V_{\rm ud} V^\ast_{\rm us}$. The Wilson coefficients $C_1$ and
$C_2$ appear in the combinations $C_2+\xi C_1$ for charged mesons
and $C_1+\xi C_2$ for neutral mesons ($\xi=1/N_c$) in the case
of the tree diagrams Ia and Ib. In the case of other diagrams including
the pole diagrams they appear in  the combinations $C_2 - C_1$.
The details of calculation of two- and three-loop quark diagrams in CCQM
may be found in our previous papers (see Refs.~\cite{Ivanov:2020xmw,
  Ivanov:2020iaq,Gutsche:2019iac,Gutsche:2019wgu,Gutsche:2018msz,
  Gutsche:2018utw,Gutsche:2017hux,Gutsche:2017wag,
  Gutsche:2015lea,Ivanov:1997ra,Ivanov:1997hi}).

Let us discuss some subtleties of calculation of the pole
diagrams.  There are contributions from  the $\frac{1}{2}^+$~resonances
(neutron and $\Sigma^+$). Note that the $\frac{3}{2}^+$~resonances 
($\Delta^0$ and $\Sigma^{\ast\,+}$) do not contribute to the amplitude
due to the KPW theorem~\cite{Korner:1970xq,Pati:1970fg}. 
This theorem states that the contraction
of the flavor-antisymmetric current-current operator with a flavor-symmetric
final state configuration is zero in the $SU(3)$ limit. The antisymmetric
$[us]$ diquark emerging from the weak vertex is in the $3^*$ representation 
and cannot evolve into the $6$ representation of the symmetric final-state 
$\{us\}$ diquark.

The propagator of the $\frac12^+$ resonances is the ordinary Dirac propagator,
\be
S(p) = \frac{1}{m_{\rm res} - \not\! p} =
\frac{m_{\rm res} + \not\! p}{m^2_{\rm res} - p^2}.
\en

Let us consider in detail the calculation of the $\Lambda\to p+\pi^-$
process which goes via neutron and $\Sigma^+$~resonances.
Recalling Eqs.~(\ref{eq:LD1}) and (\ref{eq:LD2}) one has
\be
M_{\rm LD_{1;2}} = i(2\pi)^4 \delta(p_1-p_2-q)\,
  \frac{G_F}{\sqrt{2}} V_{CKM}\, \widetilde M_{\rm LD_{1;2}}
\en
where
\bea
\widetilde M_{\rm LD_1} &=&    
- \bar u(p_2)\, \widetilde\Gamma_{n \pi p}(p_1,p_2)\, S_{n}(p_1)\,
\widetilde\Gamma_{\Lambda n}(p_1)\,u(p_1)\,,
\label{eq:neutron}\\[1.3ex]
\widetilde M_{\rm LD_2} &=&  
- \bar u(p_2)\, \widetilde\Gamma_{\Sigma^+ p}(p_2)\, S_{\Sigma^+}(p_2)\,
\widetilde\Gamma_{\Lambda \pi  \Sigma^+ }(p_1,p_2)\,u(p_1)\,.
\label{eq:sigma}
\ena
By using the Dirac equations of motion
$\bar u(p_2)\!\!\not\!p_2 = m_2\bar u(p_2)$
and $\not\!p_1 u(p_1) = m_1 u(p_1)$ one can get
\bea
&&
\bar u(p_2)\, \widetilde\Gamma_{n \pi p}(p_1,p_2) =
\bar u(p_2)\, \gamma_5\Big( C_{n \pi p} + \not\! p_1 D_{n \pi p}\Big),
\quad \widetilde\Gamma_{\Lambda n}(p_1)\,u(p_1) =
\Big( A_{\Lambda n} + \gamma_5 B_{\Lambda n}\Big) u(p_1)\,,
\nn[1.3ex]
&&
\bar u(p_2)\, \widetilde\Gamma_{\Sigma^+ p}(p_2) =
\bar u(p_2)\, \Big( A_{\Sigma^+ p} + \gamma_5 B_{\Sigma^+ p}\Big), \quad
\widetilde\Gamma_{\Lambda \pi  \Sigma^+ }(p_1,p_2)\,u(p_1) =
\gamma_5
\Big( C_{\Lambda \pi  \Sigma^+ } + \not\!p_1 D_{\Lambda \pi  \Sigma^+ }\Big)\,.
\nonumber
\ena
Finally, we arrive at the invariant matrix elements for the pole
diagrams with intermediate $\frac12^+$~resonances. One has
\bea
\widetilde M_{\rm LD_1} &\equiv& \widetilde M_{n}
= \bar u(p_2)\,\Big( A_n + \gamma_5 B_n\Big) u(p_1)\,,
\nn[1.5ex]
A_n &=& -\frac{ B_{\Lambda n} (C_{n \pi p} - m_\Lambda D_{n \pi p})}{m_n+ m_\Lambda}\,,
\qquad\qquad
B_n = -\frac{ A_{\Lambda n} (C_{n \pi p} + m_\Lambda D_{n \pi p}) }{m_n - m_\Lambda}\,,
\nn[2.5ex]
\widetilde M_{\rm LD_2} &\equiv& \widetilde M_{\Sigma}
= \bar u(p_2)\,\Big( A_\Sigma + \gamma_5 B_\Sigma \Big) u(p_1)\,,
\nn[1.5ex]
A_\Sigma &=&
-\frac{B_{\Sigma^+ p} (C_{\Lambda\pi\Sigma^+} - m_p D_{\Lambda\pi\Sigma^+} )}
{m_\Sigma+ m_p}\,,
\qquad
B_\Sigma =
-\frac{ A_{\Sigma^+ p} (C_{\Lambda \pi  \Sigma^+ } + m_p D_{\Lambda \pi  \Sigma^+ })}
{m_\Sigma - m_p}\,,
\ena
The calculation of the $A$ and $B$ amplitudes appearing in the decay
$\Lambda\to n + \pi^0$ proceeds in an analogous manner.
The pole diagram with the kaon resonance contributes only to the structure
which is proportional to $\gamma_5$. Numerically, they are negligibly small.

It is well known that the $P$-wave amplitude~$B$ is dominated by the low-lying
$1/2^+$ resonances whereas their contributions are tiny  to the $S$-wave
amplitude~$A$. The invariant amplitudes $A$ and $B$ may be converted to a set of
helicity amplitudes
$H_{\lambda_1\,\lambda_M}$ as described in~\cite{Korner:1992wi}. One has  
 \be
 H^V_{\tfrac12\,t} = \sqrt{Q_+}\,A\,,\qquad
 H^A_{\tfrac12\,t} = \sqrt{Q_-}\,B\,,
 \en 
 where $m_\pm=m_1\pm m_2$, $Q_{\pm}=m_\pm^2-q^2$.

\begin{figure}[H]
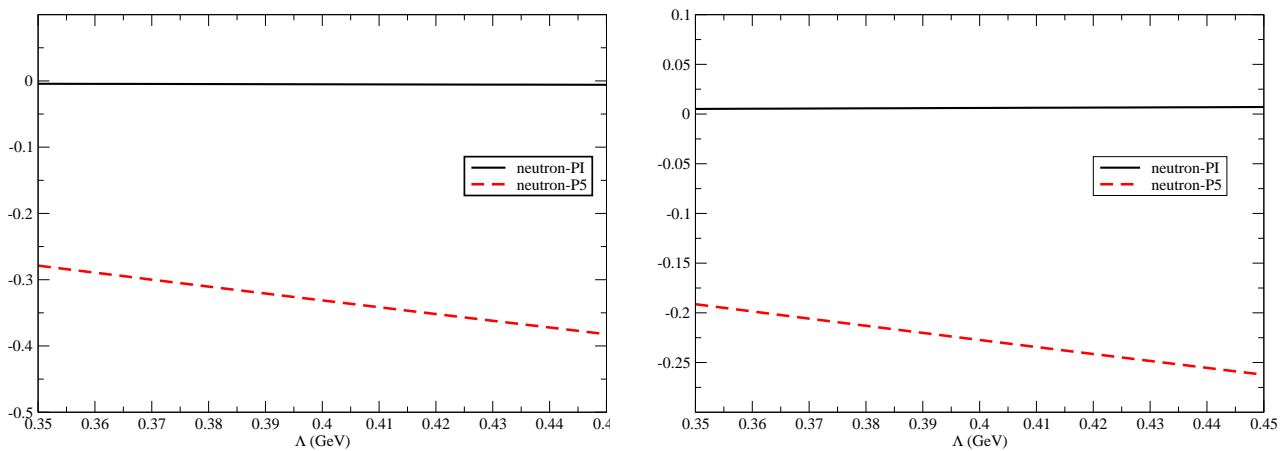

\vspace*{0.7cm}  
\begin{tabular}{lr}
\epsfig{figure=fig5a_hyperon.eps,scale=.35} \hspace*{.25cm}
\epsfig{figure=fig5b_hyperon.eps,scale=.35}
\end{tabular}
  \caption{Dependence of the helicities $PI\equiv H^V_{1/2\,t}$
    and $P5\equiv H^A_{1/2\,t}$ on the  size parameter
    in the case of the neutron resonance. Left panel: the decay
  $\Lambda\to p+\pi$; right panel: the decay  $\Lambda\to n+\pi$.}
\label{fig:neutron}
\end{figure}

 We show in Fig.~\ref{fig:neutron} the behavior of the helicity amplitudes
 as a function of the size parameter $\Lambda_s$. One can clearly see that
 the $S$-wave amplitudes are almost zero.
 
It is widely accepted that $S$-wave amplitude is saturated by the
$\frac12^-$~resonances, see, e.g., Refs.~\cite{Marshak,Bailin:1977gv,
Ebert:1983yh,Ebert:1983ih,Cheng:1985dw,Cheng:2018hwl}. Ordinarily, their
contributions are calculated by using the well-known soft-pion
theorem in the current-algebra approach. It allows one to express
the parity-violating S-wave amplitude in terms of parity-conserving
matrix elements. In our case, one has, see, e.g., \cite{Cheng:2018hwl} 
\be
A_{1/2^-}(\Lambda\to p + \pi^-) = \frac{1}{f_\pi} A_{\Lambda n}\,, \qquad
A_{1/2^-}(\Lambda\to n + \pi^0) = \frac{1}{\sqrt{2} f_\pi} A_{\Lambda n}\,,  
  \en
  where $f_\pi$ is the leptonic pion decay constant.
  Here we adopt our convention of signs.

Finally,  the transition $\frac12^+\to\frac12^+ + 0^-$ amplitude
is written in terms of invariant amplitudes as
\be
<B_2\,P|{\cal H}_{\rm eff}|B_1>
= \frac{G_F}{\sqrt{2}} V_{CKM}
\bar u(p_2)\left( A+\gamma_5\,B \right)u(p_1)\,,
\label{eq:ampl-P}
\en
where $A$ and $B$ include all relevant contributions discussed above.

The two-body decay widths read
\be
\Gamma(B_1\to B_2+P) = 
\frac{G_F^2}{32\pi}\frac{\mathbf{|p_2|}}{m_1^2}\,|V_{CKM}|^2
{\mathcal H}_S\,, \quad {\mathcal H}_S =  
  2\Big(\Big|H^V_{ \tfrac12\,t}\Big|^2 \,+\, 
  \Big|H^A_{\tfrac12\,t}\Big|^2\Big) \, 
  \label{eq:width_P}
  \en
where $\mathbf{|p_2|}=\lambda^{1/2}(m_1^2,m_2^2,q^2)/(2m_1)$.   

\section{Numerical results}
\label{sec:results}

Our covariant constituent quark model
contains a number of model parameters which have been determined by
a global fit to  a multitude of decay processes.  The values of the constituent
quark masses $m_q$ are taken from the last fit in~\cite{Gutsche:2015mxa}.
In the fit, the infrared cutoff parameter $\lambda$ of the model
has been kept fixed as found in the original paper~\cite{Branz:2009cd}.
One has
\be
\def\arraystretch{2}
\begin{array}{ccccccc}
     m_u        &      m_s        &      m_c       &     m_b & \lambda  &
\\\hline
 \ \ 0.242\ \   &  \ \ 0.428\ \   &  \ \ 1.672\ \   &  \ \ 5.046\ \   &
 \ \ 0.181\ \   & \ {\rm GeV} 
\end{array}
\label{eq: fitmas}
\en
The size parameters of light mesons were fixed by fitting the data
on the leptonic decay constants. The numerical values of the size
parameters and  the leptonic decay constants for pion and kaon are
shown in Eq.~(\ref{eq:pion}).
\be
\def\arraystretch{1.5}
\begin{array}{cccc}
 \text{Meson}        &  \Lambda_M \text{(GeV)}  & f_M  \text{(MeV)}
  &   f_M^{\rm expt} \text{(MeV)} \\
\hline
\qquad 
 \text{Pion} \qquad   &  \qquad 0.871\qquad   &  \qquad 130.3 \qquad   &
 130.0 \pm 1.7   \\
 \text{Kaon}     &   1.014   &   156.0   &   156.1 \pm 0.8   \\
\hline 
\end{array}
\label{eq:pion}
\en
In case of the nucleons, the best description of magnetic moments,
electromagnetic radii, and form factors is achieved in \cite{Gutsche:2012ze}
for a superposition of the $V–$ and $T–$currents of nucleons according
to Table~\ref{tab:cur} with $x= 0.8$ and $\Lambda_N=0.5$~GeV.
The  $\Lambda$  size parameter is the only adjustable parameter.
In Fig.~\ref{fig:Lambda} we plot the dependence on this parameter
of two branching rates $\Lambda\to p + \pi^-$ and $\Lambda\to n + \pi^0$.
One can see that the theoretical curves fit the data for
$\Lambda \approx 0.355$~GeV in both decays simultaneously.
The given value of $0.355$~GeV differs from the value of $0.492$~GeV
fixed in our previous paper \cite{Gutsche:2013pp} by fitting the experimental
data on the magnetic moment of the  $\Lambda$-hyperon.
The point is that the calculated branching fractions depend very strongly
on the  size parameter $\Lambda$ as one can see from Fig.~\ref{fig:Lambda}.
Contrary, the magnetic moment of the $\Lambda$ hyperon
calculated in the indicated paper depends very weakly on this parameter.
We have taken our old Fortran code and recalculated its value for
0.355~GeV. We found that $\mu_\Lambda=-0.74$ for 0.355~GeV, 
which is very close to the old result, $\mu_\Lambda=-0.73$ for 0.492~GeV.

For comparison, we plot the SD
contributions coming from the diagrams with topologies
Ia (charged pion), Ib (neutral pion), and  IIa, IIb, and III (all two modes).
It is readily seen that their contributions are relatively less than
those coming from the pole diagrams. However, the calculation of those
diagrams is time-consuming because it involves the analytical and
numerical calculation of three loops. The most significant contributions
among the pole diagrams are coming from the diagrams with intermediate neutron
resonance with mass closest to the $\Lambda$ resonance. The contribution
from the pole diagram with intermediate kaon resonance is negligibly small.

\begin{figure}
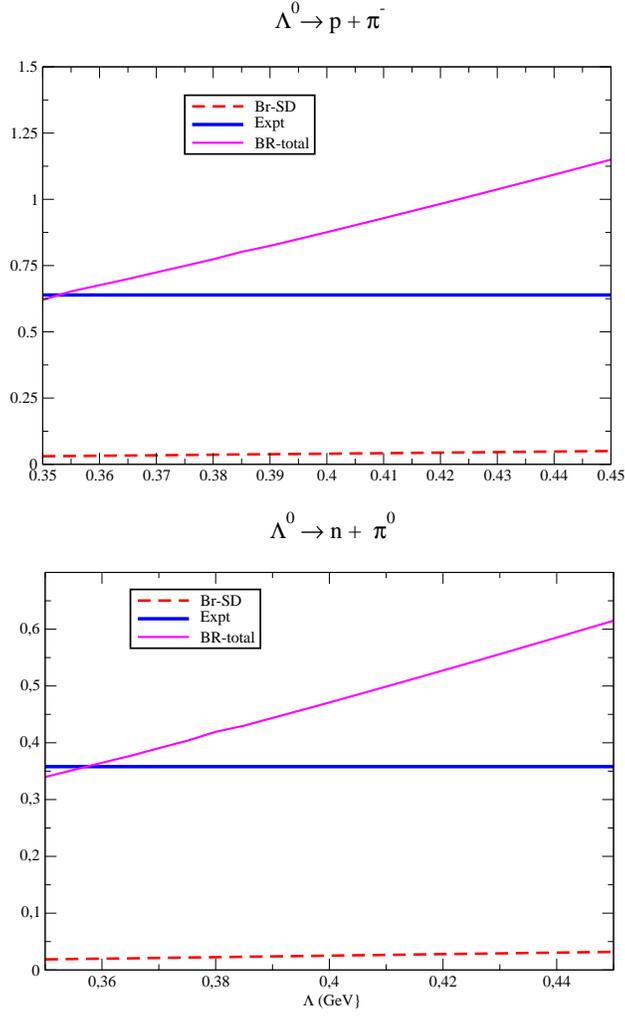

\centering
   \def\arraystretch{1}
  \begin{tabular}{c}
\epsfig{figure=fig6a_hyperon.eps,scale=.35} \\
\epsfig{figure=fig6b_hyperon.eps,scale=.35}
\end{tabular}
\caption{Dependence of the nonleptonic $\Lambda$ decay widths on
its size parameter.}
\label{fig:Lambda}
\end{figure}

In Tables~\ref{tab:SD-1}-\ref{tab:LD-2} the numerical results of 
$A_{\rm SD}$, $A_{\rm LD}$  and
$B_{\rm SD}$, $B_{\rm LD}$  at $\Lambda = 0.355$~GeV are shown.
One can see that the $B$ amplitudes dominate over $A$ amplitudes.
The SD contributions are suppressed almost by one order of magnitude comparison 
with the LD contributions.  
The numerical results for the full amplitudes are written down.
\bea
{\bf \Lambda\to p \pi^-:} && \quad A = A_{\rm SD} +  A_{\rm LD} = +0.124\,\text{GeV}^2, \quad
                                   B = B_{\rm SD} +  B_{\rm LD} = -3.042\,\text{GeV}^2.
\nn
{\bf \Lambda\to n \pi^0:} && \quad A = A_{\rm SD} +  A_{\rm LD} = +0.087\,\text{GeV}^2, \quad
                                   B = B_{\rm SD} +  B_{\rm LD} = -2.059\,\text{GeV}^2.
\label{eq:A-B}
\ena
\begin{table}[H]
  \caption{SD contibutions to the amplitudes $A$ and $B$ of the decay
    $ \Lambda\to p \pi^-$  in units of GeV$^2$.}\vskip 3mm
  \centering
  \def\arraystretch{1.3} 
  \begin{tabular}{|c|c|c|c|c|c|}
    \hline
    Amplitudes \quad 	& Ia & IIa & IIb & III & Sum\, (SD)  \\
    \hline
    $A_{\rm\, SD}$ 
    &\,\,  $-0.372 \times 10^{-1}$\quad   
    &\,\,  $ 0.269 \times 10^{-3}$\quad
    &\,\,  $ 0.300 \times 10^{-1}$\quad 
    &\,\,  $ 0.213 \times 10^{-1}$\quad 
    &\,\,  $ 0.144 \times 10^{-1}$\quad \\
    \hline
    $B_{\rm\, SD}$ & $-0.345 $ & $-0.116 $  & $ 0.167$ & $ -0.452 $
    & $ -0.746 $ \\
    \hline
  \end{tabular}
  \label{tab:SD-1}
\end{table}	
\begin{table}[H]
  \caption{LD contibutions to the amplitudes $A$ and $B$ of the decay
    $ \Lambda\to p \pi^-$  in units of GeV$^2$.}\vskip 3mm
  \centering
  \def\arraystretch{1.3} 
  \begin{tabular}{|c|c|c|c|c|c|c|}
    \hline
    Amplitudes \quad 	& $n$ & $\Sigma^+$ & $K$ & $K^\ast$ & $\tfrac12^-$\, (Soft pion) &
    Sum\, (LD)  \\
    \hline
    $A_{\rm\, LD}$ 
    &\,\, $-2.13 \times 10^{-3}$\quad 
    &\,\, $-9.54 \times 10^{-3}$\quad
    &\,\, $0$ \quad 
    &\,\, $2.61  \times 10^{-2}$\quad 
    &\,\, $0.869 \times 10^{-1}$\quad  
    &\,\, $1.10  \times 10^{-1}$\quad\\
    \hline
    $B_{\rm\, LD}$ &\,\, $-2.55 $\quad 
    &\,\, $2.26 \times 10^{-1}$\quad
    &\,\, $2.82 \times 10^{-2}$\quad  
    &\,\, $0$ \quad & \,\, $0$\quad 
    &\,\, $-2.296 $\quad \\
    \hline
  \end{tabular}
  \label{tab:LD-1}
\end{table}
%
%
\begin{table}[H]
  \caption{SD contibutions to the amplitudes $A$ and $B$ of the decay
    $ \Lambda\to n \pi^0$  in units of GeV$^2$.}\vskip 3mm
  \centering
  \def\arraystretch{1.3} 
  \begin{tabular}{|c|c|c|c|c|c|}
    \hline
    Amplitudes \quad 	& Ib & IIa & IIb & III & Sum\, (SD)  \\
    \hline
    $A_{\rm\, SD}$ 
    &\,\,  $-0.120 \times 10^{-1}$\quad   
    &\,\,  $ 0.190 \times 10^{-3}$\quad
    &\,\,  $ 0.211 \times 10^{-1}$\quad 
    &\,\,  $ 0.150 \times 10^{-1}$\quad 
    &\,\,  $ 0.243 \times10^{-1}$ \quad \\
    \hline
    $B_{\rm\, SD}$ & $-0.112 $ 
    & $-0.82\cdot 10^{-1} $  
    & $ 0.119$ 
    & $ -0.319 $
    & $ -0.394 $ \\
    \hline
  \end{tabular}
  \label{tab:SD-2}
\end{table}
\begin{table}[H]
  \caption{LD contibutions to the amplitudes $A$ and $B$ of the decay
    $ \Lambda\to n \pi^0$ in units of GeV$^2$.}\vskip 3mm
  \centering
  \def\arraystretch{1.3} 
  \begin{tabular}{|c|c|c|c|c|c|c|}
    \hline
    Amplitudes \quad 	& $n$ & $\Sigma^0$ & $K$ & $K^\ast$ & $\tfrac12^-$\, (Soft pion) &
    Sum\, (LD)  \\
    \hline
    $A_{\rm\, LD}$ 
    &\,\, $-1.52 \cdot 10^{-3}$\quad 
    &\,\, $-6.58 \cdot 10^{-3}$\quad
    &\,\, $0$ \quad 
    &\,\, $8.44  \cdot 10^{-3} $ \quad 
    &\,\, $6.24  \cdot 10^{-2} $\quad  
    &\,\, $0.627 \cdot 10^{-1} $\quad\\
    \hline
    $B_{\rm\, LD}$ 
    &\,\, $-1.83 $\quad  
    &\,\, $1.56  \cdot 10^{-1}$\quad
    &\,\, $0.902 \cdot 10^{-2}$\quad 
    &\,\, $0$ \quad  
    &\,\, $0$ \quad 
    &\,\, $-1.665 $\quad \\
    \hline
  \end{tabular}
  \label{tab:LD-2}
\end{table}

\section{Summary and conclusion}
\label{sec:summary}

We have studied two-body nonleptonic decays of light lambda 
hyperon $\Lambda \to p \pi^- (n\pi^0)$  
with account for both short- and long-distance effects. 
The short-distance effects are induced 
by five topologies of external and internal weak $W$ interactions, 
while long-distance effects are saturated by an inclusion of 
the so-called pole diagrams. Pole diagrams are generated by resonance
contributions of the low-lying spin $\frac12^+$ (nucleon and $\Sigma$)
and spin $\frac12^-$ baryons. The last contributions are calculated
by using the well-known soft-pion theorem. The spin $\frac32^+$~resonances
do not contribute to the amplitude due to the K\"orner-Pati-Woo theorem.
The contributions from the intermediate $K$~meson is also negligibly  small.
From our previous analysis of heavy baryons we have known that 
short-distance effects induced by internal topologies are not 
suppressed in comparison with external $W$-exchange diagrams 
and must be included for description of data. 
Here, in the case of $\Lambda$ decays we have found that the contribution 
of the SD diagrams is sizably suppressed, almost 
by one order of magnitude in comparison with data, which are 
known with quite good accuracy.  
The most significant contributions
are coming from the intermediate $\frac12^+$ and $\frac12^-$ resonances.
The contribution from the kaon resonance is negligibly small.

   
\begin{acknowledgments}

This work was funded 
by BMBF ``Verbundprojekt No. 05P2018 - Ausbau von ALICE
am LHC: Jets und partonische Struktur von Kernen''
(F\"orderkennzeichen No. 05P18VTCA1), 
by ANID PIA/APOYO AFB180002,  
by FONDECYT (Chile) under Grant No. 1191103, 
and by Millennium Institute for Subatomic Physics
at the High-Energy Frontier (SAPHIR) of ANID, Code: ICN2019\_044 (Chile).  
M.A.I.\ acknowledges the support from the PRISMA Cluster of Excellence
(Mainz Uni.). M.A.I. and J.G.K. acknowledge the Heisenberg-Landau Grant for
partial support.
Zh.T.'s research has been funded by the Science Committee of the Ministry of
Education and Science of the Republic of Kazakhstan (Grant No. AP09057862).

\end{acknowledgments}  


\end{document}